# Production of nuclear sources and nuclear batteries by proton irradiation

S. Möller*, T. Wegener

*Forschungszentrum Jülich GmbH, Institut für Energie- und Klimaforschung – Plasmaphysik, 52425 Jülich, Germany*

The decay of instable nuclei is being used in a broad range of applications from detector calibration to power sources. As the public acceptance of classical fission nuclear technology is decaying and its integral costs are enormous, alternative and flexible production routes are required. The physical formalism and fundamental considerations are presented for the use of ion accelerators for isotope production.

22 isotopes are selected from the nuclide chart offering the best properties for different applications. A focus is put on the production of nuclear power sources to substitute the scarce Pu-238 based batteries. For power sources 20 MeV protons are found to produce α emitting polonium isotopes from bismuth with the highest energy efficiency of 0.031%. 14kC are required to produce a $1W_{th}$ power source of the 2.9 year half-life α emitter Po-208. Within some month nuclear power sources for space missions can be produced in a suitable accelerator. The technical requirements are close to and compatible with the planned International Fusion Materials Irradiation Facility accelerator offering a cost-effective dual use perspective.

Keywords: Nuclear battery, radioisotopes, cyclotrons, accelerators, space propulsion, nuclear sources

## I. INTRODUCTION

Nuclear batteries, also known as atomic batteries, are widely used in special applications where long battery life is required and costs are of minor importance. Applications include the use in space, medicine and remote areas. Nuclear isotopes used in these batteries can be natural ones like uranium, but are mainly products of nuclear fission reactors due to their shorter half-lifes and thus higher power density. Currently one of the main isotopes is Pu-238 as it offers easy handling by being an α emitter and its 88 years half-life are long enough for the relevant applications and short enough to provide good output power per nucleus. In the future fusion reactors and accelerators may serve as valuable tools to add isotopes to this list. For α emitters typically the nuclear energy is converted to heat which is then converted to electricity, e.g. by thermo-elements (Blanke, Birden, Jordan, & Murphy, 1960). In general also a direct conversion of nuclear to electric energy is possible by semiconductors, e.g. so-called beta-voltaic (Olsen, 1973). Having a broad range of possible emitters is thus favourable.

The problem with the artificial production of isotopes is the required irradiation dose to obtain a relevant power output from the battery. As a back-of-an-envelope calculation a release of 1 MeV=$1.6*10^{-13}$ J per decay and an efficiency of the production of electricity from the decay of 1% an activity of $10^{15}$ Bq is required for 1.6 W of electrical output. Modern energy conversion techniques or the direct use of heat e.g. for space propulsion could improve that number, but the general problem of required high activities remains due to the nature of the involved processes. In a fission reactor these activities can be produced by the high neutron fluxes. NASA is frequently using kW range Pu-238 batteries for space missions, but as reactors are being shut-down serious supply issues are foreseen (Witze, 2014).

The method presented in this work is based on using industrial standard accelerators, especially cyclotrons, to induce the required activity. As the accelerator technology and also the knowledge of ion-material interaction and nuclear cross-sections have significantly improved in the last decades this method becomes attractive. With energies of about 30 MeV and ion currents of >1 mA for proton beams, the production of nuclear power sources e.g. by (p, n) reactions becomes feasible from basically any isotope using state-of-the-art accelerators. No nuclear chemistry is required for separation and also basically no unwanted isotopes are produced by this method, leading to significantly lower nuclear waste production compared to fission reactors. Ion irradiation is already widely used for the production of positron emitters for medical applications, but the technology has the potential to deliver any radioactive product a fission reactor can produce. The method allows also for the production of nuclear power sources at mass and, due to the higher flexibility of ion irradiation, spatially and time-wise, closer to the final use. By adjusting the ion fluence on the material, the output power can be varied by orders of magnitude from pW to kW making them applicable to miniature devices as well as propulsion systems. For example a space operated cyclotron could serve as a kind of gas station by producing the required power sources for a flight to Mars directly in the near earth region, thus eliminating dangers of loss of nuclear material during transportation from earth to space.

## II. NUCLEAR ACCUMULATOR DESIGN

### A. Charging/Activity calculation

In the context of ion beam production of a nuclear battery the term accumulator seems more adequate due to a number of differences to neutron irradiation based sources. A nuclear accumulator is per-se not producing (significant) nuclear decay heat, only by the irradiation its activity is realized: It is being charged. The accumulator is also not a primary source of energy, but energy is being stored in the accumulator. This energy has to be invested from outside, i.e. by the power put into an accelerator. In this work charging by MeV protons is considered, without loss of generality. The protons can be provided e.g. by a cyclotron, which is a reliable and cost efficient device for providing some 10 MeV proton beams. For the calculation of the charging process the reaction cross-sections and the stopping power of protons in the corresponding primary material are required. All technical efficiencies are assumed to be 100% in the calculations. The cross-sections $\sigma(E)$ are taken from the nuclear code Talys 1.8 and the TENDL-2015 database (Koning, et al., 2015), as it offers all the required cross-sections and the code is generally seen to reproduce experimental production cross-sections within a factor of about 2, as can be seen e.g. in the JANIS4 database (Soppera, Bossant, & Dupont, 2014). For most of the considered reactions, experimental data are available and the uncertainties are usually seen to be much smaller. Stopping powers $S(E)[Energy/Length]$ are derived from the Bethe-Bloch equation resulting in deviations in the order of 5% from SRIM 2013 values (Ziegler, Biersack, & Littmark, 2013) in the considered energy range of 5-30 MeV.

The main equation for the activity calculation of a nuclear accumulator is

$$A = N_T * I_{beam} * H * (1 - e^{-\frac{t_{irr}}{\tau}}) * \int_0^{E_0} \sigma(E)/S(E) dE \qquad (1)$$

with the number of target atoms $N_T$, the abundance of the relevant isotope $H$, the ion current $I_{Beam}$. The exponential term represents the decay with time constant $\tau$ during loading time $t_{irr}$ and the integral gives the reaction probability of each projectile. It is assumed that the accumulator consists of a pure and un-charged material with $H=1$. The equation applies to each possible reaction by summing the cross-sections of all reactions leading to a certain product.

The choice of 30 MeV protons becomes obvious when looking at the values of $\sigma(E)$. In most cases a reaction threshold can be found in the 5-10 MeV region (negative Q-value), followed by a maximum in cross-section of about 0.1-1 barn at 10-20 MeV and a slight decrease to higher energies (Soppera, Bossant, & Dupont, 2014). At the same time the stopping power decreases to higher energies with stopping power maxima being at about 100 keV. As the ratio of these values is relevant for an efficient charging of the accumulator an optimal point in ion energy is somewhere around 30 MeV. Besides this 30 MeV is selected for simplicity and comparison reasons, but slightly different energies might be more efficient for a certain isotope. The gain in efficiency by further optimization of the beam energy will probably be less than a factor of 3, which is small compared to the benefit of selecting the right isotopes, and is neglected here due to the unknown systematic uncertainties of TENDL cross-sections. The use of α particles and (α, xn) reactions is also possible, but due to the significantly higher stopping powers of α particles their production efficiency will be significantly smaller than with hydrogen isotopes (see equation 1). Theoretical cross-sections are found to be comparable to (p, xn) reactions (Koning, et al., 2015), though.

The specific output $P$ gives a number for the amount of present atoms of the radioactive isotope required to produce 1 W of output power

$$P = (\frac{\ln(2)}{\tau} * e^{\frac{\ln(2)}{\tau}} * E_{decay} * e)^{-1} \qquad (2)$$

where $E_{decay}$ is the energy per nuclear decay of the isotope. Together with the reaction probability the irradiation dose in units of ion charge (fluence) per W of output power can be calculated

$$A = \frac{P*e}{\int_0^{E_0} \sigma(E)/S(E)\,dE} \tag{3}$$

The efficiency of the accumulator is then given by the beam energy $E_{beam}$ and the stored energy per ion

$$\eta = \frac{E_{decay}}{E_{beam}} * \int_0^{E_0} \frac{\sigma(E)}{S(E)}\,dE \tag{4}$$

## B. Nuclide selection

The selection of the primary accumulator nuclide is the first step in designing a nuclear accumulator. The nuclide is selected to fulfill the following requirements:

- Product can be reached from a stable isotope by a (p,n) or (p,2n) reaction
- Reaction cross-section is high and the stopping power is low in order to reduce irradiation time and maximize charging efficiency
- Half-life of the product is suitable for the planned application: Long enough for the required minimum energy output at the end of use, but as short as possible for a minimum production time (eq. 3). As the lower limit a half-life of 1 year was chosen as below this time-frame chemical batteries are favorable
- Maximum energy released per decay for minimum irradiation time and maximum efficiency
- Emitted radiation can be effectively converted to energy (α and β emitters favorable, high energy γ emitters are unfavorable)

Besides that the material has to fit the application requirements e.g. regarding melting temperature or chemical properties, but this is not considered here. The list of feasible nuclides becomes rather short compared to the list of possible nuclides with this amount of requirements. For half-lifes of about 0.6-680 years the optimal nuclides are listed below in table 1. From the list of these nuclides several can be eliminated as many of them are in the 1 year half-life range. Favorable candidates are selected according to the requirements above and highlighted in dark grey for the different ranges of half-lifes in view of a low proton irradiation time/high efficiency and low gamma energies. Nuclides in light grey have a single interesting property but lack in fulfilling other requirements.

The total efficiency of the nuclear accumulators is very low as only about 1-5 MeV are released per decay, but 30 MeV are invested in the production of a nucleus (the rest is lost by stopping) and in addition only <1% of the protons actually induce a nuclear reaction. In conclusion, the input energy to the accelerator will be about $10^5$ times higher than what is

Production of nuclear sources and nuclear batteries by proton irradiation released by the nuclear accumulator during its operation. If the further conversion to electricity is taken into account, up to $10^{-6}$ total electricity efficiency can be provided.

Many of the produced nuclides are β+/ec (Electron Capture) instable products. In this case the exhausted accumulator will mainly consist of the original material and can be recharged at any time by further irradiation. This is not true for all isotopes as the existence of other proton induced reactions as e.g. (p, xn), (p, 2p), (p, α) lead to additional change of the material composition not considered here. Also, the production of neutrons and their subsequent reaction with the material is not considered. This might be less important for the output power, but can be relevant for radiation dose rates or standard sources.

Table 1: Selection chart of the possible materials for a nuclear accumulator and their corresponding products induced by proton irradiation. The favourable materials are highlighted for the different half-life ranges of 0.6 to 680 years. For the evaluation (p,n) and (p,2n) reactions are considered as these offer the highest cross-sections in the 30MeV range of protons. Reactions are assumed to produce the ground state nuclei. In case several primary isotopes can be used for production the one with the highest cross-section is taken and highlighted. Nuclear data are extracted from ENDF/B-VII.1 using (Nucleonica GmbH, 2014), cross-sections are taken from TENDL-2015. All isotopes are assumed to be produced instantly (with infinite beam current). Calculations for Bi-209 based on 20MeV protons. The relevant uncertainties in the calculated quantities are related to uncertainties of the production cross-sections which can only be estimated as stated in section 2.1.

| Primary nuclides | Product | Half-life [years] | Mean gamma energy [MeV] | Energy per decay [MeV] | Specific output power [Atoms/W] | Production probability ∫ σ/S dE | Irradiation dose per thermal output [C/W] | Efficiency at 30 or 20* MeV |
|---|---|---|---|---|---|---|---|---|
| Ne-22 | Na-22 | 2.60 | 2.20 | 2.80 | 2.64E+20 | 0.33% | 1.28E+04 | 3.08E-04 |
| Cr-54 | Mn-54 | 0.85 | 0.84 | 1.38 | 1.75E+20 | 0.33% | 8.49E+03 | 1.52E-04 |
| Fe-57, Fe-58 | Co-57 | 0.75 | 0.13 | 0.84 | 2.54E+20 | 0.20% | 2.03E+04 | 5.60E-05 |
| Cu-65 | Zn-65 | 0.67 | 0.58 | 1.35 | 1.41E+20 | 0.23% | 9.82E+03 | 1.04E-04 |
| Zr-91.92 | Nb-91 | 680.00 | 0.01 | 1.25 | 1.55E+23 | 0.20% | 1.24E+07 | 8.33E-05 |
| Ru-101.102 | Rh-101 | 3.30 | 0.29 | 0.54 | 1.74E+21 | 0.15% | 1.85E+05 | 2.70E-05 |
| Ru-102 | Rh-102 | 2.90 | 2.12 | 2.32 | 3.55E+20 | 0.14% | 4.06E+04 | 1.08E-04 |
| Xe-134 | Cs-134 | 2.07 | 1.55 | 2.06 | 2.86E+20 | 0.01% | 4.57E+05 | 6.87E-06 |
| Cs-133 | Ba-133 | 10.54 | 0.40 | 0.52 | 5.76E+21 | 0.08% | 1.15E+06 | 1.39E-05 |
| Nd-144.145.146 | Pm-144 | 1.00 | 1.56 | 2.33 | 1.22E+20 | 0.29% | 6.73E+03 | 2.25E-04 |
| Nd-145.146 | Pm-145 | 17.70 | 0.03 | 0.16 | 3.15E+22 | 0.30% | 1.68E+06 | 1.60E-05 |
| Nd-146 | Pm-146 | 5.53 | 0.75 | 1.50 | 1.05E+21 | 0.05% | 3.35E+05 | 2.50E-05 |
| Sm-150 | Eu-150 | 36.40 | 1.53 | 2.30 | 4.50E+21 | 0.05% | 1.44E+06 | 3.83E-05 |
| Sm-152 | Eu-152 | 13.53 | 1.16 | 1.85 | 2.08E+21 | 0.01% | 3.33E+06 | 6.17E-06 |
| Eu-153 | Gd-153 | 240.40 | 0.10 | 0.48 | 1.42E+23 | 0.04% | 5.70E+07 | 6.40E-06 |
| Gd-157.158 | Tb-157 | 99.00 | 0.01 | 0.06 | 4.69E+23 | 0.30% | 2.50E+07 | 6.00E-06 |
| Gd-158 | Tb-158 | 180.00 | 0.80 | 1.17 | 4.37E+22 | 0.04% | 1.75E+07 | 1.56E-05 |
| Yb-173.174 | Lu-173 | 1.34 | 0.17 | 0.67 | 5.69E+20 | 0.27% | 3.37E+04 | 6.03E-05 |

| Yb-174 | Lu-174 | 3.60 | 0.17 | 1.40 | 7.31E+20 | 0.04% | 2.92E+05 | 1.87E-05 |
| Pb-207 | Bi-207 | 31.80 | 1.50 | 2.40 | 3.77E+21 | 0.05% | 1.21E+06 | 4.00E-05 |
| Bi-209 | Po-209 | 102.00 | 0.005 | 5.00 (α) | 5.80E+21 | 0.05% | 1.86E+06 | 8.33E-05 * |
| Bi-209 | Po-208 | 2.93 | 2E-5 | 5.20 (α) | 1.59E+20 | 0.18% | 1.41E+04 | 3.12E-04 * |

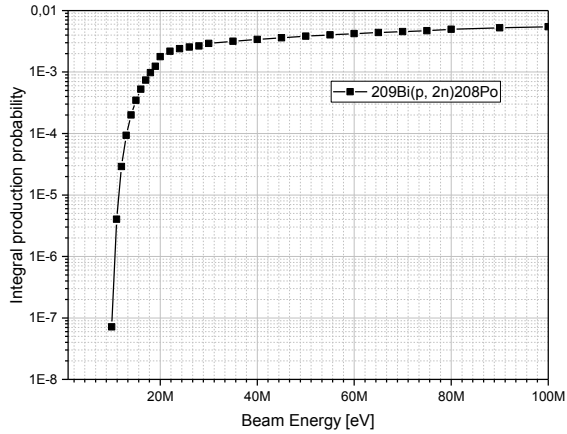

Figure 1: The production probability of Po-208 on irradiation of Bi-209 by protons and its relation to the proton beam energy. At around 20 MeV the production efficiency starts to saturate at some 0.1%.

## C. Bismuth and Zirconium

The use of Bismuth differs from the other primary materials as α emitters are produced. α emitters are favorable for a nuclear accumulator as the range of the decay products is small and low dose rates can be expected for the surroundings. The Bismuth accumulator can include up to 3 different relevant isotopes in the selected energy range coming from (p, 1-3n) reactions. The existence of these isotopes can be controlled via the beam energy, as the reactions have negative Q-value. For this purpose, the proton energy should be kept below 20MeV in order to avoid the production of Po-207 which will quickly decay to Bi-207, both producing high energy gammas and thus dose rate problems (Nucleonica GmbH, 2014). The decay of Po-208 produces Pb-204 which is a quasi-stable isotope. Po-209 produces Pb-205 which releases up to 85 keV photons with a half-life of 17.3 My. For this reason, the shielding of the decay products radiation is also uncritical. The applied cross-sections are backed by experimental data leading to an estimated uncertainty of 30% in the production probability. From the cost side Bi has the advantage of being a cheap and isotopically pure primary material which can also be oxidized to form a stable oxide ceramic or it can be used as a direct voltage source in a $Bi_2Te_3$ thermoelectric generator. The direct irradiation of $Bi_2Te_3$ is feasible, as the main isotopes with relevant half-lifes produced from Te are I-129 (<40 keV γ) and I-125 (150 keV ec). A separation of Bi and Po during and after production is possible via electrolytic, chemical or distillation processes. An in-situ separation of Bi and Po during isotope production using hot liquid Bi as target and separating the Po at a condenser at <962°C could provide a cost efficient method for removing the beam heat while keeping the target isotopically pure. The energy efficiency of production by using Bi is the highest of all investigated nuclides and supports 20 MeV as good irradiation energy in this case, see figure 1. Bismuths low melting point and thermal conductivity may lead to the requirement for a robust capsule of higher melting temperature during high power charging or when used for thermal space propulsion with hot hydrogen gas. Alternatively Po can be used as oxide in a power source, offering much higher melting temperatures.

Zr-91 and Zr-92 are interesting for producing the longest lived isotope (Nb-91, electron capture decay) which also has mainly low gamma energies. Zirconium has low neutron absorption and another stable isotope (Zr-90) Therefore even from secondary neutron reactions no significant activity is expected. For very long lasting applications or when a constant source is required Nb-91 offers the best performance among the investigated products. For all other applications it is recommended to irradiate Bi to get Po products because of the 3 orders of magnitude lower specific irradiation dose required.

### III. CONCLUSIONS

The proposed technology for the production of nuclear accumulators by ion irradiation remains a costly, but probably competitive product for the use in special applications. The power efficiency is calculated to be at maximum 0.031%. The advantages over fission produced sources are a gain in flexibility, as targets, beam energy and irradiation dose can be varied at each point in time. Furthermore no nuclear chemistry is required and less nuclear waste is produced as the instable isotopes are directly produced in a matrix of stable isotopes and all nuclear reaction products are produced intendedly. The method is based on currently available standard accelerators as cyclotrons or linear accelerators. These devices are frequently used for the production of medical isotopes from gases, but water-cooled targets for solid samples are also available (Spellerberg, et al., 2015). New target concepts optimized for the nuclear accumulator production as the liquid bismuth distillation target could be required for maximum production rates at ion beam currents in the range of mA to A. At these high production rates, constant isotope separation, e.g. by electrolytic, chemical or distillation means, becomes necessary to maintain pure targets.

The production requires a high current accelerator producing 20-40 MeV hydrogen ions. In this context "high" is defined as a current allowing producing the desired activity in a time much smaller than the half-life of the desired isotope, see equation (1). A possible selection scheme of the isotopes for a nuclear accumulator is presented. Within this scheme the selection is a trade-off between half-life and output power, low gamma dose rate vs. high energy release per decay (efficiency) and general material properties and availability. Bismuth is identified as the most interesting primary material for the production of nuclear power sources. It has the advantages of being abundant and having only a single stable isotope and a relatively high efficiency of the power storage. As the products of the (p, n) and (p, 2n) reactions Po-208 and Po-209 are α emitters of about 5 MeV, low gamma dose rates, comparable to those of Pu-238 batteries, are expected for the environment. Some additional activity of Po-210 might be induced by the neutrons released from the proton reaction, depending on the technical design of the irradiation system. $1.6*10^{20}$ nuclear reactions (produced Po-208 atoms) are required for 1 W of output power. With an accelerator delivering 1mA of proton current this results in about 5 month of continuous irradiation time. For producing kW Po-208 accumulators an ion current ≥1 A is required. Technical concepts providing these numbers are already designed for example in the context of the neutron source IFMIF (Mosnier, Beauvais, Branas, & al, 2010). The neutrons produced in the (p, xn) reactions used for Po-208 production could provide a dual use of the system, improving cost efficiency for both sides.

The electric energy cost for a Po-208 accumulator of 1 $W_{th}$ initial output is 14 kC*20 MeV=78 MWh divided by the efficiency of the accelerator and plus the expenses for cooling, vacuum and control systems. This number is quite substantial and shows the need for efficient accelerator systems, but also evaluation of improved production mechanisms using ion channelling or deuterons and the importance of a secondary gain by exploiting the neutrons coming from the (p, xn) reactions for other applications.

The use of nuclear accumulators in space propulsion is a possible field of application. Here the risks of sending radioactive materials into space are significant and not generally accepted. The possibility to produce the nuclear battery in space in an orbital or moon based cyclotron may be a crucial advantage for long-term trips as a Mars mission. Furthermore in this application nuclear batteries offer a simultaneous source of heat, electrical energy and propulsion power. The accumulator can be encapsulated, e.g. in a thick tungsten housing, in order to provide high temperatures and shielding at the same time. The high temperature power source can then be used for propulsion with hot hydrogen or methane gas. The half-life of Po-208 is much shorter than that of Pu-238, but a space or moon based cyclotron could serve as a filling station for the space ships.

With the range of protons in the 30 MeV range being in the order of mm, isotopes can also be directly produced in or implanted/drifted into a semiconductor for the production of small scale highly integrated nuclear batteries. In the same way isotope sources for calibration and analysis purposes can be produced in the range of Bq to TBq of activities by varying beam current and exposure time. A large set of nuclides is available to fit basically all needs for gamma, neutron, electron, positron and alpha energies, similar to fission produced products. The isotopes will be different from the known fission produced isotopes though, as mostly (p, n) reactions are applied in contrast to (n, γ) dominating the isotope production in fission reactors.

All considerations were, so far, based on the use of protons. Deuterons may provide a possibly more efficient alternative to protons with higher cross-sections for the (d, xn) reactions compared to (p, xn). For the production of Po-208 TENDL proposes an increased cross-section of 1.75 times. Unfortunately the experimental knowledge of the required reaction cross-sections is even more limited than for protons, requiring more research for a credible assessment of the application of deuterons.

<div align="center">IV. REFERENCES</div>